\documentclass[fleqn,10pt]{wlscirep}
\usepackage[utf8]{inputenc}
\usepackage[T1]{fontenc}
\usepackage{amsmath}

\title{Excitonic Effects in Absorption Spectra of Carbon Dioxide Reduction Photocatalysts}

\author[1]{Tathagata Biswas}
\author[1,*]{Arunima K. Singh}

\affil[1]{Department of Physics, Arizona State University, Tempe, AZ, 85287}
\affil[*]{arunimasingh@asu.edu}

\begin{abstract}
The formation and disassociation of excitons plays a crucial role in any photovoltaic or photocatalytic application. However, excitonic effects are seldom considered in materials discovery studies due to the monumental computational cost associated with the examination of these properties. Here, we study the excitonic properties of nearly 50 photocatalysts using state-of-the-art Bethe-Salpeter formalism. These $\sim$50 materials were recently recognized as promising photocatalysts for CO$_2$ reduction through a data-driven screening of 68,860 materials. Here, we propose three screening criteria based on the optical properties of these materials, taking excitonic effects into account, to further down select 6 materials. Remarkably we find a strong correlation between the exciton binding energies obtained from the Bethe-Salpeter formalism and those obtained from the computationally much less-expensive Wannier-Mott model for these chemically diverse $\sim$50 materials. This work presents a new paradigm towards the inclusion of excitonic effects in future materials discovery for solar-energy harvesting applications.
\end{abstract}

\begin{document}

\flushbottom
\maketitle

\thispagestyle{empty}

\section*{Introduction}

Excitons, which are quasiparticles, form when light is absorbed by semiconductors or insulators. They consist of a photoexcited electron and a hole bound by Coulomb interaction. In photocatalytic and photovoltaic processes these excitons must dissociate into free carriers and eventually reach the reacting species (in a photocatalytic process) or get converted into photocurrent (in a photovoltaic device) before any recombination processes can occur. Spontaneous generation of electron-hole (e-h) pairs from the dissociation of excitons is only possible when exciton binding energy (EBE) is lower than $T \Delta S_{diss}$, where $T$ is the environmental temperature and $\Delta S_{diss}$ is the entropy increase upon dissociation \cite{giebink2011thermodynamic}. Once free e-h pairs are available, a low carrier recombination rate is desirable as it ensures most of these e-h pairs can be utilized in photocatalytic/photovoltaic applications. Both radiative, which are predominant in direct gap semiconductors, as well as non-radiative recombination processes have adverse effects on the quantum efficiency of a photocatalytic/photovoltaic device. However, with the help of innovative solutions such as use of multi-junction photocatalysts or solar cells \cite{meng2018tio2,dimroth2007high} one can go beyond the limit set by radiative recombination processes. In the context of solar-cells, this limit is given by the Shockley–Queisser limit \cite{shockley1961detailed} and a lower limit is expected for photocatalysts \cite{hashemi2014solar}. Non-radiative recombinations that fall in the category of Shockley-Read-Hall \cite{shockley1952statistics} or trap-assisted recombinations can also be controlled by improving device quality and using defect-free semiconductors \cite{richter2013reassessment}. The other kind of non-radiative recombination, known as Auger recombination \cite{auger1923rayons,beattie1959auger} is quite hard to eliminate \cite{vossier2010auger}. High exciton binding energy is detrimental for a photovoltaic/photocatalytic device as it can facilitate these Auger recombination processes \cite{beattie1959auger,yang2015comparison}. Therefore, it is important to consider excitonic effects while looking for promising photovoltaics and photocatalysts.
 
Excitonic effects have been seldom considered in material discovery studies conducted in the past for identifying promising photovoltaic or photocatalytic materials \cite{wang2017insights,guo2016mxene,shinde2017discovery,yan2017solar}. For instance, recently, Singh et. al. \cite{singh2019robust} performed the largest CO$_2$ photocathode search to date, where they shortlisted 52 materials starting from a list of 68,860 candidate materials. Their first-principles computation-based screening strategy was based on evaluating the thermodynamic stability, the electrochemical stability,~\cite{singh2017electrochemical} and the electronic structure compatibility with the CO$_2$ reduction reaction\cite{singh2019robust}. Their screening criteria were highly selective of robust photocatalysts when applied to the scores of known photocatalysts. However, despite its success, the screening did not incorporate any excitonic effects. In recent years, Bethe-Salpeter formalism (BSE) has gained remarkable success in probing excitonic effects in bulk \cite{umari2014relativistic,landmann2012electronic,schmidt2019quasiparticle} as well as 2-dimensional materials \cite{tran2014layer,ugeda2014giant,qiu2013optical}, where these effects are more pronounced. The BSE formalism has a high computing resource requirement, both in terms of CPU time and memory. But, it has been found to have excellent agreement with experimental measurements across various materials classes. \cite{umari2014relativistic,landmann2012electronic,schmidt2019quasiparticle, tran2014layer,ugeda2014giant,qiu2013optical}

In this study, we design a materials screening strategy which is based on the optical properties and considers excitonic effects to identify the most efficient photocatalysts for CO$_2$ reduction among the 52 materials shortlisted by Singh et. al. We employ many-body perturbation theory within GW approximation together with \cite{hedin1965new,hybertsen1986electron,shishkin2007self} BSE to explore quasiparticle and optical properties. Our screening strategy involves three distinct properties \textemdash EBE of the materials, the capability to absorb incident (visible or UV) radiation, and the degree of anisotropy in absorbing light of different polarizations. For each of these properties, we propose the limiting value for the criteria by looking at some of the well-known photocatalysts used for CO$_2$ reduction. Using these criteria we identify a shortlist of 6 suitable materials from Singh et al's study that display optical properties which are at par or better than known photocatalysts. This screening approach is broadly applicable to the discovery of materials for other solar-energy applications. The limiting value corresponding to each of the properties in such a screening strategy will, however, change based on the specific application. 

As the BSE is computationally resource intensive and time-consuming, we also assess the prospect of using the computationally much less demanding Wannier-Mott model to compute the EBE of materials. We compare the EBE obtained from the BSE method and the Wannier-Mott model for 48 materials from Singh et al's study. These materials are chemically diverse \textemdash comprising of oxides, sulphides, tellurides, and arsenides. We find a strong correlation between the EBE obtained from the BSE and that obtained from the Wannier-Mott model. Thus, we surmise that the Wannier-Mott model can be used in place of the BSE method to estimate the EBE of materials in a computational cheaper, hence potentially a high-throughput manner, to enable computational data-driven materials discovery for applications involving solar-absorption such as photovoltaics and photocatalysis.   

\section*{Results and Discussion}

\subsection*{GW-BSE vs HSE (Do we need GW-BSE?)}

Singh et al. \cite{singh2019robust} had employed density-functional theory (DFT) with a hybrid exchange-correlation functional to compute the bandgaps and band edges of potential photocatalysts and identified materials that can utilize the visible-light spectrum which accounts for 44\% of the solar radiation and simultaneously facilitate the reduction reaction of CO$_2$. DFT which is designed to explore ground state properties suffers from bandgap underestimation problem as the bandgap is an excited state property of materials. Hybrid Functionals such as HSE and HSE06, overcome this issue by adding a fraction of Hartree-Fock exchange to the traditional exchange-correlation potential (LDA or GGA) and is quite successful in predicting correct bandgaps of a wide variety of materials. While hybrid functionals typically exhibit improved treatment of semiconductor bandgaps, they are heavily reliant on the fraction of the exact exchange included in the functional and are thus limited in their predictive capacity. This limitation is especially severe in the case of certain groups of materials where strong correlation effects are important, such as, transition metal oxides \cite{vines2017systematic}. 

Many-body perturbation theory within GW approximation is often an order of magnitude more expensive than standard DFT with hybrid functionals but doesn't require any empirical parameters and is found to be remarkably successful in case of various materials classes. The GW formalism combined with BSE has been proven to be the state-of-the-art method to study the electronic structure and optical properties of a wide variety of semiconductor and insulator materials in recent years with an excellent predictive capacity. \cite{onida2002electronic,malone2013quasiparticle,PhysRevB.89.075205} Moreover, the GW-BSE computed absorption spectra have a very good agreement with experimentally measured spectra. Thus, we reevaluate the bandgaps of 48 photocatalysts from Singh et al's study using the GW method. We have restricted our study to the non-magnetic 48 of the 52 materials identified by Singh et al. We exclude the ZrVF$_6$ and the two phases of VOF as they have a magnetic ground state and the CdHgAsBr since it was found to be metallic according to our GW calculations.  

\begin{figure}[!h]
\centering
\includegraphics[width=\linewidth]{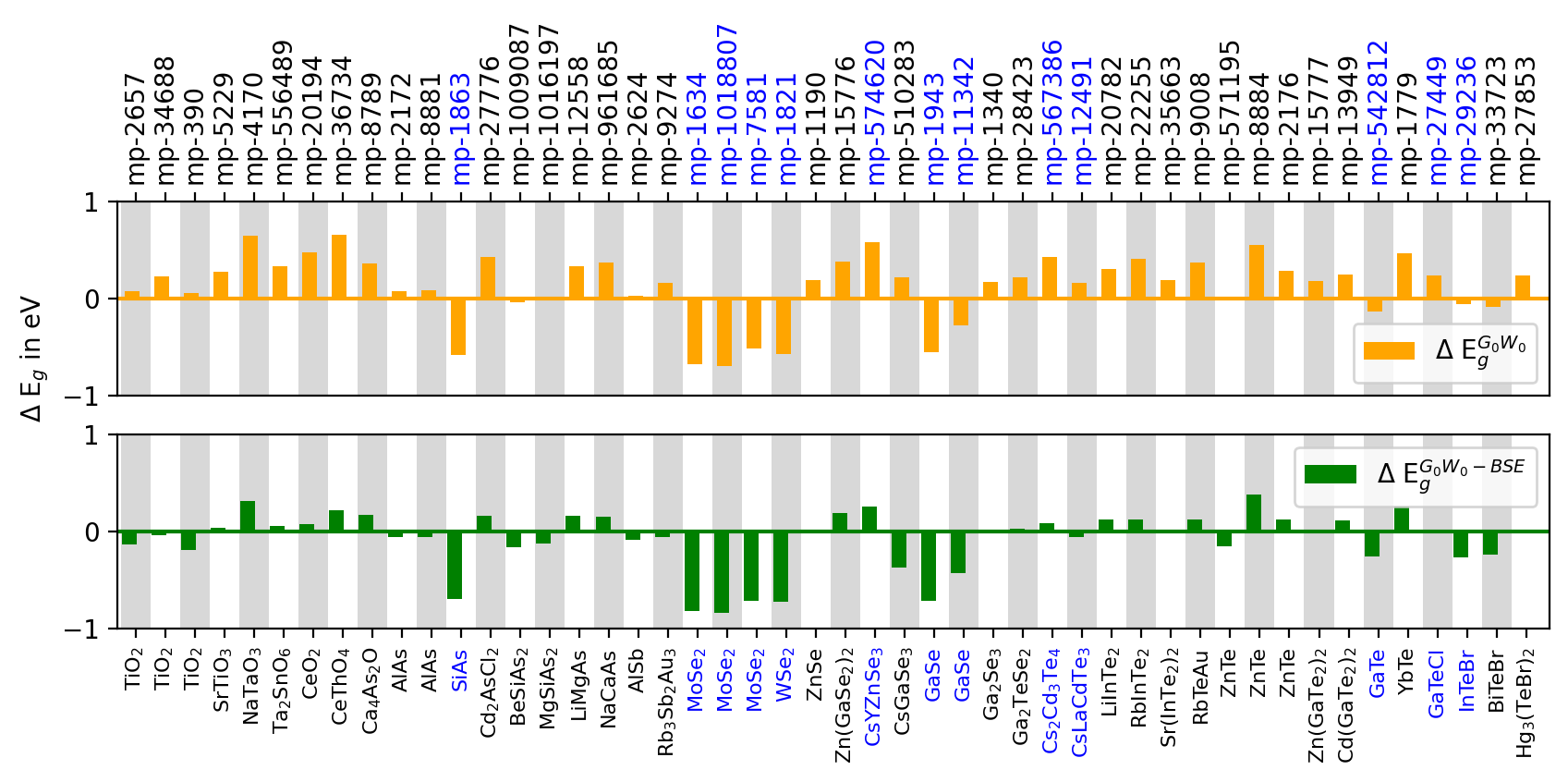}
\caption{Differences between $E_\mathrm{g}^\mathrm{HSE}$ and
	$E_\mathrm{g}^\mathrm{G_0W_0}$ (orange) as well as
	$E_\mathrm{g}^\mathrm{G_0W_0-BSE}$ (green). The Materials Project id's for each material noted on the bottom $x$-axis are listed on the top $x$-axis ticks. The layered materials have been labeled in blue.}
\label{fig2}
\end{figure}

Fig. ~\ref{fig2} shows the difference between the HSE06 bandgaps,
$E_\mathrm{g}^\mathrm{HSE}$, reported in Singh et al's work and the quasiparticle (QP) gaps obtained from
G$_0$W$_0$, $E_\mathrm{g}^\mathrm{G_0W_0}$, as well as the optical gaps
obtained from G$_0$W$_0$-BSE, $E_\mathrm{g}^\mathrm{G_0W_0-BSE}$ calculations. The difference between the quasiparticle gap of a material and its optical gap is precisely its EBE. One should note that, G$_0$W$_0$ refers to non-self-consistent GW approximation. Here, the single-particle Green's function, G$_0$, and the screened Coulomb interaction, W$_0$, are calculated using Kohn-Sham eigenvalues and eigenfunctions \cite{shishkin2006implementation}. 

We have calculated a root mean
square difference (RMSD), defined as $\sqrt{\frac{1}{N}\Sigma^N_{i=1} 
(\Delta{E_\mathrm{g,i})^2}}$, for the gaps obtained from G$_0$W$_0$ and GW$_0$-BSE with the HSE bandgap values for all the 48 materials. Where $E_\mathrm{g,i}$ is the bandgap of the $i^\mathrm{th}$ of the $N$=48 materials considered in this study. Note that a positive $\Delta{E_\mathrm{g,i}}$ signifies underestimated HSE band
gap. From Fig. ~\ref{fig2} it is evident that except the layered materials,
the HSE bandgap is almost always significantly underestimated compared to the
QP gap obtained from G$_0$W$_0$ calculations. We find an RMSD value of 0.16 eV for optical gaps and
0.31 eV for QP gaps for the non-layered materials considered in this study,
which highlights the importance of studying photocatalytic materials with
the GW-BSE methodology. In the case of layered materials, the difference between previously reported $E_\mathrm{g}^\mathrm{HSE}$ and $E_\mathrm{g}^\mathrm{G_0W_0-BSE}$ resulting from our calculation, has a different origin, which we will discuss in the following section.

%As the QP gaps are computed by including self-energy correction within
%G$_0$W$_0$ approximation to the Kohn-Sham (KS) eigenvalues, they are expected
%to be larger than the Kohn-Sham bandgaps, as can be seen in SI. Table 2. The
%optical gaps were computed by finding the lowest energy excitation from the
%solution of BSE. The difference between the direct QP gap and the optical gap
%is precisely the binding energy of an exciton. The indirect absorption edges
%reported in this work were estimated by subtracting the EBE from the QP gap of
%the indirect bandgap materials. We emphasize here that, at finite temperatures
%the optical gaps obtained by solving the BSE directly corresponds to
%experimental absorption edges in the case of direct bandgap materials. Since,
%in case of indirect gap materials the light absorption for photons with the
%energy less than the direct bandgap has to be phonon assisted and we haven't
%included any electron-phonon interactions in our calculations.Thus we only
%report only the estimated absorption edges.  

\subsection*{Layered materials and Van der Waals correction}

\begin{table}[!h]
\centering
\footnotesize
\begin{tabular}{ | l| l| l| l| l| l| l| l| l| l| l| l| l| l| l|}
\hline
mp-id  &   Formula  &   SG  &  $a^\mathrm{Theo}$ &  $b^\mathrm{Theo}$ & $c^\mathrm{Theo}$  &  $\alpha^\mathrm{Theo}$ &   $\beta^\mathrm{Theo}$ &  $\gamma^\mathrm{Theo}$ &  $a^\mathrm{Expt}$ &  $b^\mathrm{Expt}$ & $c^\mathrm{Expt}$  &  $\alpha^\mathrm{Expt}$ &   $\beta^\mathrm{Expt}$ &  $\gamma^\mathrm{Expt}$ \\
\hline
\hline
mp-1863           &   SiAs              &   C2/m              &          16.15  &       3.72   &         9.64  &         90.00  &       105.76    &        90.00   &        15.98   &         3.67   &         9.53   &        90.00   &       106.00   &        90.00  \\
\hline
mp-1943           &   GaSe              &   P6$_3$/mmc          &         3.80   &         3.80   &        15.73   &        90.00   &        90.00   &       120.00   &         3.75   &         3.75   &        15.99   &        90.00   &        90.00   &       120.00  \\
\hline
mp-11342          &   GaSe              &   R3m               &         3.80   &         3.80   &        24.14   &        90.00   &        90.00   &       120.00  &         3.75   &         3.75   &        23.91   &        90.00   &        90.00   &       120.00  \\
\hline
mp-542812         &   GaTe              &   C2/m              &         17.61   &         4.14   &        10.62   &       90.00   &        104.60   &       90.00   &        17.40   &         4.08   &        10.46   &        90.00   &       104.50   &        90.00  \\
\hline
mp-1634           &   MoSe$_2$          &   P6$_3$/mmc          &         3.33   &         3.33   &        13.17   &        90.00   &        90.00   &       120.00   &         3.29   &         3.29   &        12.93   &        90.00   &        90.00   &       120.00  \\
\hline
mp-1018807        &   MoSe$_2$          &   P6$_3$/mmc          &         3.32   &         3.32   &        13.06   &        90.00   &        90.00   &       120.00   &         3.28   &         3.28   &        12.92   &        90.00   &        90.00   &       120.00  \\
\hline
mp-7581           &   MoSe$_2$          &   R3m               &         3.33   &         3.33   &         19.78   &        90.00   &        90.00   &       120.00   &         3.29   &         3.29   &        19.39   &        90.00   &        90.00   &       120.00  \\
\hline
mp-1821           &   WSe$_2$           &   P6$_3$/mmc          &         3.31   &         3.31   &        13.04   &        90.00   &        90.00   &       120.00   &         3.28   &         3.28   &        12.96   &        90.00   &        90.00   &       120.00  \\
\hline
mp-27449          &   GaTeCl            &   Pnnm              &         5.90   &         14.59   &        4.13   &        90.00   &        90.00   &        90.00   &         5.85   &        14.47   &         4.08   &        90.00   &        90.00   &        90.00  \\
\hline
mp-29236          &   InTeBr            &   P2$_1$/c            &         7.46   &         7.70   &         8.32   &        90.00   &       115.45   &        90.00   &         7.35   &         7.58   &         8.34   &        90.00   &       117.61   &        90.00  \\
\hline
mp-569454         &   CdHgAsBr          &   Pmma              &         5.17   &         8.93   &         9.80   &        90.00   &        90.00   &        90.00   &         4.70   &         8.79   &         9.78   &        90.00   &        90.00   &        90.00  \\
\hline
mp-574620         &   CsYZnSe$_3$       &   Cmcm              &         4.14   &         15.82   &        10.93   &        90.00   &        90.00   &       90.00   &         4.14   &        15.81   &        10.93   &        90.00   &        90.00   &        90.00  \\
\hline
mp-567386         &   Cs$_2$Cd$_3$Te$_4$   &   Ibam              &         6.74   &        13.16   &        15.56   &        90.00   &        90.00   &        90.00   &         6.67   &        13.01   &        15.39   &        90.00   &        90.00   &        90.00  \\
\hline
mp-12491          &   CsLaCdTe$_3$      &   Cmcm              &         4.66   &         17.32   &        12.17   &        90.00   &        90.00   &       105.60   &         4.64   &        16.70   &        12.17   &        90.00   &        90.00   &        90.00  \\
\hline
\end{tabular}
\caption{Comparison between the computed (Theo) and experimental (Expt) lattice
	constants ($a$, $b$, $c$ in \AA, and $\alpha$, $\beta$, $\gamma$ in
	$^{\circ}$), the Materials Project material ids (mp-id) and spacegroup (SG) of  all the
	layered materials considered in this study. The experimental lattice constants are derived from the ICSD~\cite{belsky2002new} as reported in the Materials Project database.} 
\label{tab2}
\end{table}

We observe a large difference in the previously reported
$E_\mathrm{g}^\mathrm{HSE}$ values and our $E_\mathrm{g}^\mathrm{G_0W_0}$ as
well as $E_\mathrm{g}^\mathrm{G_0W_0-BSE}$ values for the layered materials. This
difference originates primarily from the change in the lattice parameters due
to the Van der Waals (vdW) corrections included in our work. Singh et. al. used the lattice
parameters computed by the Materials Project for their computational screening of more than 68,000 potential candidates\cite{singh2019robust}. The Materials Project structural relaxation simulations do not account for any vdW corrections. For the layered materials, we have included van der Waals (vdW) corrections using the non-local van der Waals density functional (vdW-DF) proposed by Dion et al \cite{dion2004van} with optimized exchange functionals (optB88) \cite{klimevs2009chemical}. 

In Table.~\ref{tab2} we compare the lattice parameters obtained by using vdW corrections for the structural optimization in case of the layered materials with the experimental lattice parameters and we find a very good agreement. The experimental lattice constants are obtained from the ICSD~\cite{belsky2002new} lattice constants available through the Materials Project database. We also find that for these layered materials the Materials Project computed lattice parameters, and hence those used in Singh et al's work, are typically underestimated in the direction of the vdW bonding. This gives rise to larger $E_\mathrm{g}^\mathrm{HSE}$ bandgaps in Singh et al's work and the significantly smaller $E_\mathrm{g}^\mathrm{G_0W_0}$ and  $E_\mathrm{g}^\mathrm{G_0W_0-BSE}$ gaps in this work. A RMSD of 0.54 eV for optical gaps and 0.47 eV for QP gaps is observed for the layered materials considered in this study, underscoring the importance of vdW corrections. See Supplementary Table S1 for the lattice parameters of all the 48 materials obtained from the vdW corrected functional. 

In the case of the direct bandgap layered materials, GaSe and GaTe, the optical gap is found to be smaller than 1.7 eV, the lower end the of desirable 1.7 eV to 3.5 eV bandgap window of a visible-light photoabsober.~\cite{singh2019robust}For the dichalcogenides MoSe$_2$ (all three phases) and  WSe$_2$ the direct absorption edge is just above 1.5 eV, however, at finite temperature the phonon assisted indirect absorption edge will be at an energy lower than 1.5 eV. These observations suggest that these materials may not be ideal for visible-light photocatalysis due to their low optical gap.

\subsection*{G$_0$W$_0$ vs GW$_0$ (Which one is better?)}

We also computed the bandgaps using the partially self-consistent GW$_0$ and GW$_0$-BSE calculations. In GW$_0$, the eigenfunctions from ground-state calculations (i.e. DFT) are maintained, and G$_0$W$_0$ calculations are performed to compute QP energies, which is then updated in the calculation of the Green's function, G \cite{shishkin2007self}. These calculations show a significant increase in the QP gap for most of the oxides as well as several other materials such as CsYZnSe$_3$ and RbTeAu (see Supplementary Information Table S2). As a result, the RMSD value for the QP gap obtained from the GW$_0$ calculation shoots up to 0.8 eV. A relatively low RMSD value of 0.32 eV is obtained for the optical gap, as the EBE's are larger ($\sim$ 0.5 eV) for these materials. 

\begin{table}[!h]
\centering
\begin{tabular}{|l|l|l|l|l|}
\hline
Material & $E_\mathrm{g}^\mathrm{G_0W_0-BSE}$ & $E_\mathrm{g}^{\mathrm{GW_0-BSE}}$& Experimental \\
\hline
\hline
TiO$_2$ (rutile) & 3.07 & 3.91 & 3.03 \cite{amtout1995optical}\\
\hline
TiO$_2$ (anatase) & 3.31 & 4.19 & 3.42 \cite{tang1995urbach}\\
\hline
SrTiO$_3$ & 3.24 & 4.21 & 3.30 \cite{tezuka1994photoemission} \\
\hline
CeO$_2$ &  3.68 & 6.04 & 3.78 \cite{griffiths1976spectroscopic} \\
\hline
MoSe$_2$ & 1.08 & 1.14 & 1.09 \cite{kam1982detailed}\\
\hline
WSe$_2$ & 1.18 & 1.25 & 1.20 \cite{kam1982detailed}\\
\hline
ZnSe & 2.30 & 2.65 & 2.82 \cite{passler1999temperature} \\
\hline
ZnTe & 2.22 & 2.49 & 2.39 \cite{passler1999temperature} \\
\hline
AlAs & 2.04 & 2.26 & 2.15 \cite{madelung2012semiconductors}\\
\hline
AlSb & 1.71 & 1.89 & 1.62 \cite{madelung2012semiconductors}\\
\hline
\end{tabular}
\caption{Comparison between optical gaps obtained from G$_0$W$_0$-BSE, $E_\mathrm{g}^\mathrm{G_0W_0-BSE}$, 
	GW$_0$-BSE calculation, $E_\mathrm{g}^{\mathrm{GW_0-BSE}}$, and the experimental values previously reported
	in the literature. All values are in eV. The
	$E_\mathrm{g}^\mathrm{G_0W_0-BSE}$ is in better agreement with the
	experimentally measured values in comparison to the
	$E_\mathrm{g}^{\mathrm{GW_0-BSE}}$.} \label{tab1}
\end{table}

Table~\ref{tab1} presents a comparison of the $E_\mathrm{g}^\mathrm{G_0W_0-BSE}$ and $E_\mathrm{g}^\mathrm{GW_0-BSE}$ with experimentally measured optical gaps. Experimental measurements of the optical gaps were available in the literature for 10 of the materials that we considered in this work -- four oxides, three selenides, one arsenide, and an antimonide. The $E_\mathrm{g}^\mathrm{G_0W_0-BSE}$ has a better agreement (RMSD=0.19 eV) with the experimentally obtained optical gaps than the $E_\mathrm{g}^\mathrm{GW_0-BSE}$ (RMSD=0.86 eV). This might be due to the fact that the four oxides are wide bandgap materials and self-consistency tends to overestimate the bandgap in such materials \cite{van2006quasiparticle}. The only materials for which the GW$_0$-BSE provides better agreement with experiments is ZnSe and ZnTe. In case of MoSe$_2$ and WSe$_2$ self-consistency doesn't change the QP gap significantly and both G$_0$W$_0$-BSE and GW$_0$-BSE provide excellent agreement with experiments.
Therefore, in all the subsequent calculations reported in this study we have restricted ourselves with G$_0$W$_0$ level of calculation to find the QP energies.

\subsection*{Design of the Screening Criteria}

We designed three screening criteria to identify the photocatalysts with the most suitable optical properties among the 48 materials considered in this study. Below we describe these criteria in detail. 

\subsubsection*{Criteria 1: Exciton Binding Energy} 
As we have already discussed, high EBE is not desirable for photocatalytic or photovoltaic applications. Therefore, we consider materials that have an EBE of less than 200 meV as suitable candidates for photocatalysis. The motivation behind choosing 200 meV as the EBE criteria limit came from looking at the experimental EBE of materials that are currently accepted as a
viable choice for photocatalytic applications, such as the anatase phase of TiO$_2$ (EBE=180 meV)
\cite{baldini2017strongly} and MoS$_2$ (EBE=240 meV) \cite{park2018direct}. 
We find that 26 of the 48 materials considered in this study have an EBE of less than 200 meV. We examine these materials in detail in the Outcome of Screening for Optical Properties section.

\subsubsection*{Criteria 2: Average Integrated Absorption Coefficient} 

While it is necessary for a photoabsorber or photocatalytic material to have an optical gap in the visible light region to absorb a significant amount of sunlight, it is not a sufficient condition. It is also necessary that the material is able to absorb visible light of all possible polarizations. 
The amount of incident radiation of a certain frequency and polarization that will be absorbed by a material is determined by its absorption coefficient at that particular frequency and polarization axis. 

Considering this, we propose to use an additional screening 
criteria to search for good photoabsorber/photocatalytic materials, one that is  based on their
integrated absorption coefficient, $\alpha_\mathrm{int}$, for light polarization along the x ($\parallel$ $\hat{a}$), y ($\parallel$
$\hat{b}$) and z ($\parallel$ $\hat{c}$) crystal axes, namely $\alpha_\mathrm{int}^{x}$, $\alpha_\mathrm{int}^{y}$, and $\alpha_\mathrm{int}^{z}$. 

$\alpha_\mathrm{int}$ is computed by integrating the frequency
dependent absorption coefficient ($\alpha(\omega)$) in the 1.7-3.5 eV range\cite{singh2019robust} for
a polarization of light along x ($\parallel$ $\hat{a}$), y ($\parallel$
$\hat{b}$) and z ($\parallel$ $\hat{c}$) axes. To compute $\alpha(\omega)$ one
needs both the real ($\epsilon_1(\omega)$) and the imaginary
($\epsilon_2(\omega)$) part of the dielectric function. In case of light polarization along x,

\begin{equation}
\alpha_\mathrm{int}^x = \int_{\omega_{min}}^{\omega_{max}}{\alpha^x(\omega)} d\omega \\ \mathrm{where} \hspace{8pt}
\alpha^x(\omega)=\frac{2 \pi \lvert \sqrt{(\epsilon^x_1(\omega))^2+ (\epsilon^x_2(\omega))^2} \rvert - \epsilon^x_1(\omega)} {\lambda} 
\end{equation}
where $\omega$ and $\lambda$ are the frequency and wavelength of incident radiation.
Here, we obtain $\epsilon_2(\omega)$ directly by solving BSE and use the
Kramers–Kronig relation to calculate $\epsilon_1(\omega)$.

Through this integration of the $\alpha(\omega)$ in the energy range of visible
photons and averaging for different light  polarization, one can eliminate all
the materials that have excitons in the desired energy range but are optically
inactive or `dark'. Optically `dark' excitons exist as a solution to the BSE
but have zero oscillator strength, hence, do not contribute to optical
absorption. 

To find a suitable cutoff value of $\alpha^\mathrm{avg}_\mathrm{int}$ (defined as $\frac{\alpha^{x}_\mathrm{int}+\alpha^{y}_\mathrm{int}+\alpha^{z}_\mathrm{int}}{3}$) for identifying promising photocatalysts we again look at the most widely studied material for this application, the anatase phase of TiO$_2$. However, the anatase TiO$_2$ is not used as a visible-light photocatalyst. The optimum performance for this material can only be achieved under exposure to UV radiation. Therefore, we
have computed the $\alpha^\mathrm{avg}_\mathrm{int}$ in the range of 3.5--4.2 eV for TiO$_2$
and used that as the cutoff. We find the value of $\alpha^\mathrm{avg}_\mathrm{int}$ for
TiO$_2$ is 3.6  $\times10^4$ cm$^{-1}$ eV. This value is almost three times smaller than the $\alpha^\mathrm{avg}_\mathrm{int}$ for silicon (10.5 $\times10^4$ cm$^{-1}$ eV), the most widely used photoabsorber of this time. The $\alpha^\mathrm{avg}_\mathrm{int}$ for
silicon was computed using experimental data for $\alpha(\omega)$ \cite{Green1995} in the
visible range of 1.7--3.5 eV. 

\subsubsection*{Criteria 3: Anisotropy in Integrated Absorption Coefficient} 

Anisotropic absorption of light is undesirable as it would make certain surfaces of a poly-crystalline
sample incapable of absorbing incident non-polarized solar radiation, and hence,
reducing the efficiency of the solar-to-chemical energy conversion process. Thus, we consider an additional screening criteria, $\alpha^\mathrm{aniso}_\mathrm{int}$, that assess the anisotropy in the integrated absorption coefficients along $x$, $y$, and $z$ polarization axis. Here,  $\alpha^\mathrm{aniso}_\mathrm{int} = \frac{minimum(\alpha^{x}_\mathrm{int}, \alpha^{y}_\mathrm{int}, \alpha^{z}_\mathrm{int})}{maximum(\alpha^{x}_\mathrm{int}, \alpha^{y}_\mathrm{int}, \alpha^{z}_\mathrm{int})}$. By computing $\alpha^\mathrm{aniso}_\mathrm{int}$ one can eliminate materials that absorb lights with preferential polarization. We chose an $\alpha^\mathrm{aniso}_\mathrm{int}$ of > 0.8 as the criteria of selection for this study. 

In summary, the three criteria that we propose for screening photocatalysts with suitable optical properties are,

\begin{enumerate}
	\item EBE < 200 meV
	\item $\alpha^\mathrm{avg}_\mathrm{int}$ >  3.6 $\times10^4$ cm$^{-1}$ eV 
	\item $\alpha^\mathrm{aniso}_\mathrm{int}$ >  0.8 
\end{enumerate}

\subsection*{Outcome of Screening for Optical Properties}

\subsubsection*{Exciton Binding Energy}

\begin{figure}[!h]
\centering
\includegraphics[width=0.75\linewidth]{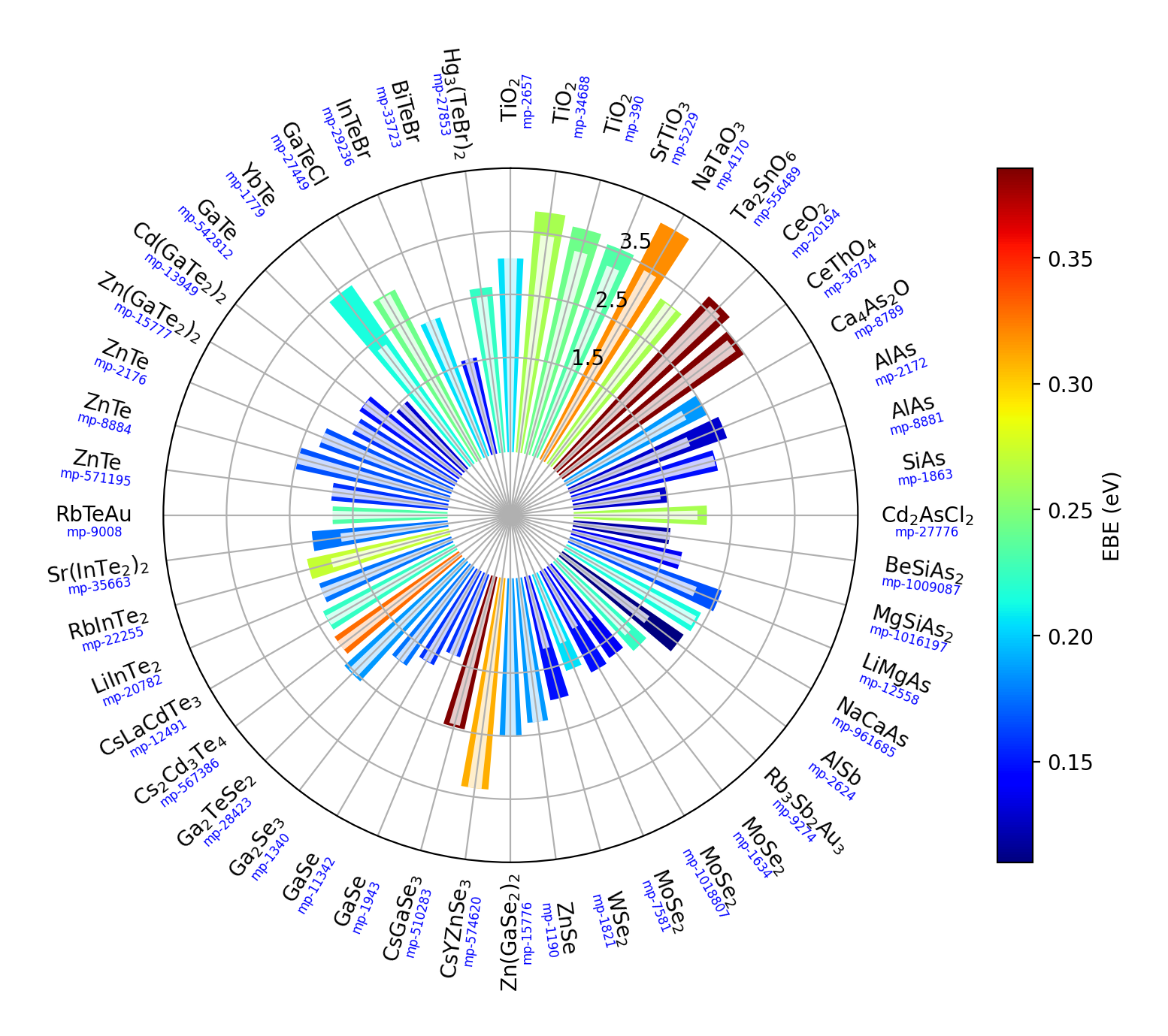}
\caption{The bar heights denote the optical gap, $E_\mathrm{g}^\mathrm{G_0W_0-BSE}$ in eV, and their colors denote the
	exciton binding energy (EBE) as obtained from G$_0$W$_0$-BSE calculations for the 48 materials. We find that 26 of the 48 materials have EBE less than 0.2 eV. The radial heights corresponding to 1.5 eV, 2.5 eV, and 3.5
	eV optical gaps are labeled in the figure. The indirect absorption edges are
	shown as white bars for each material. For direct gap materials, they
	are equal to the optical gap.} \label{fig1}
\end{figure}

In Fig. ~\ref{fig1} we present the calculated optical gaps, $E_\mathrm{g}^\mathrm{G_0W_0-BSE}$, and the EBE of all the 48 materials. The bar heights denote the $E_\mathrm{g}^\mathrm{G_0W_0-BSE}$ and their colors denote the exciton binding energy (EBE). Most of the oxides considered here have high EBE's, making them unappealing for photovoltaic and photocatalytic applications. We, however, find
26 materials with very low EBE's of $<$ 200 meV. The EBEs of all the 48 materials computed from both G$_0$W$_0$-BSE as well as GW$_0$-BSE are listed in the Supplementary Table S2. 

With the exception of NaCaAs, all the materials considered in this study that have a F$\bar{4}$3m spacegroup
(AlSb, AlAs, ZnTe, ZnSe, and LiMgAs) have EBE $<$ 200 meV. All of these materials have a Zinc-blend crystal structure except LiMgAs, which has an additional Mg atom
at the center of the unit cell. AlAs and ZnTe with Wurtzite crystal structure (spacegroup P6$_3$mc) also have quite low EBE's, 145 meV and 169
meV respectively. Among the materials with low EBE ($<$ 150 meV) we find two group-II–IV–V$_2$ ternary compounds with body centered tetragonal chalcopyrite structure
(spacegroup I$\bar{4}$2d) namely, MgSiAs$_2$ and BeSiAs$_2$. These two materials are known to be promising materials for spintronics, electronic and optoelectronic applications such as infrared-nonlinear optical material, solar energy converters, infrared detectors, visible and invisible
light-emitting diodes \cite{chiker2011birefringence,cheddadi2017first}. 

A group of telluride materials (Zn(GaTe$_2$)$_2$, Cd(GaTe$_2$)$_2$,
Sr(InTe$_2$)$_2$), which are generally synthesized by preparing solid solutions
of A$_2$Te$_3$ (A= Ga, In) with other A$^{\mathrm{II}}$B$^{\mathrm{IV}}$
tellurides,\cite{woolley1960effects} also exhibit low EBE's (140--180 meV). The layered chalcogenides (GaTe, GaSe, MoSe$_2$, and WSe$_2$) have EBE $<$ 200 meV as well. Interestingly, the different phases of these chalcogenides do not have a
significant effect on the EBE of these materials. Most of the layered
dichalcogenides (MoSe$_2$, WSe$_2$) with hexagonal crystal unit cell
(spacegroup P6$_3$/mmc) have EBE in the range of (130--150 meV). MoSe$_2$ in a hypothetical phase with rhombohedral unit cell (spacegroup R3m),
that has a slightly larger EBE of 208 meV.

\subsubsection*{Integrated Absorption Coefficient and Its Anisotropy}

\begin{figure}[h!]
\centering
\includegraphics[width=\linewidth]{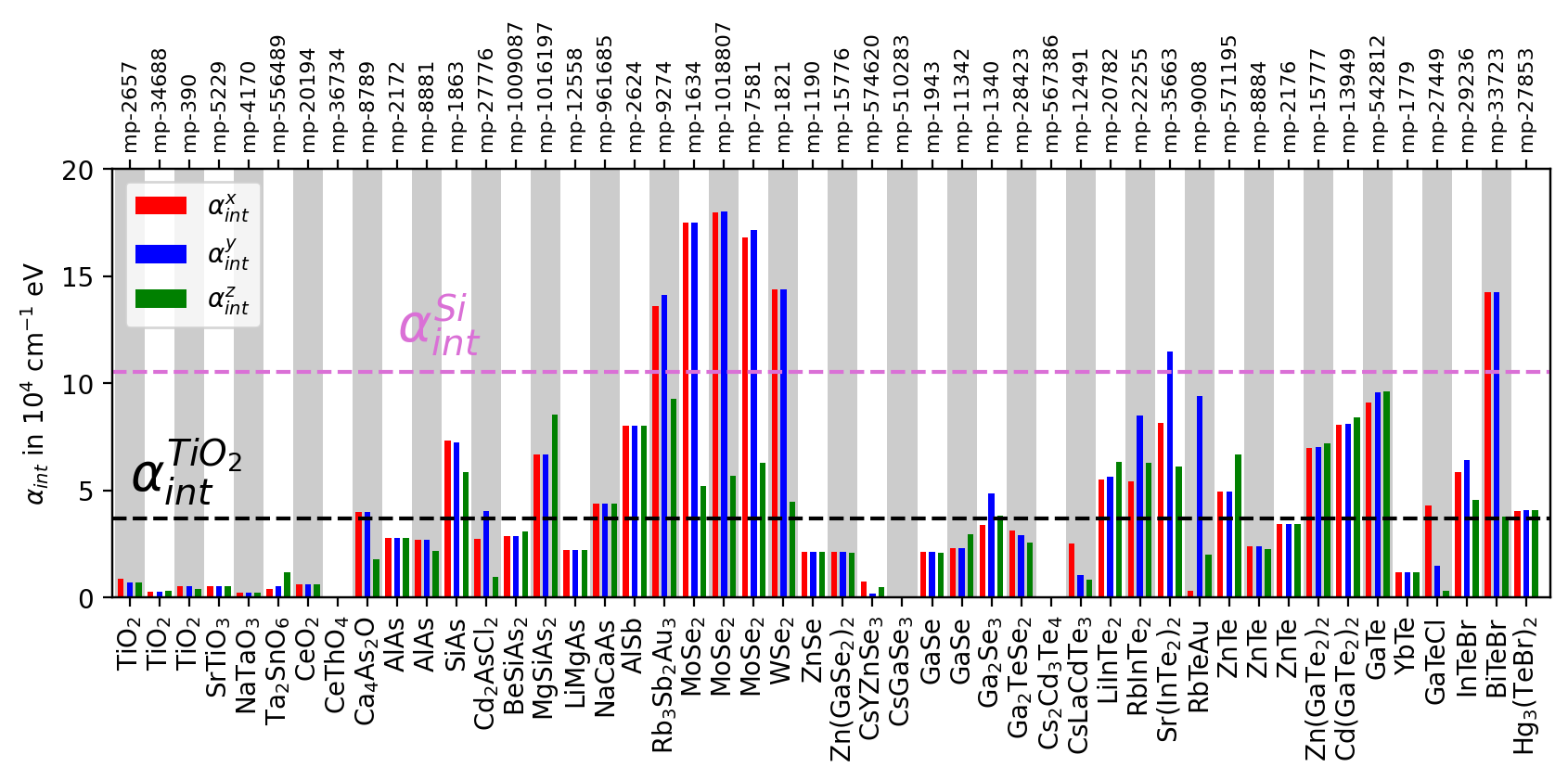}
\caption{Integrated absorption coefficient ($\alpha_\mathrm{int}$) in case of light
	polarization along $x$ (red bars), $y$ (blue bars), and $z$ (green bars) axis for all the
	materials under consideration. We have integrated the absorption
	coefficient over the 1.7-3.5 eV energy range to show the degree of
	anisotropy in absorption for the visible region of the solar spectrum.
	Horizontal dashed lines denote the $\alpha_\mathrm{int}$ values for silicon
	(magenta) and TiO$_2$ (black).} 
\label{fig3}
\end{figure}

\begin{figure}[ht]
\centering
\includegraphics[width=\linewidth]{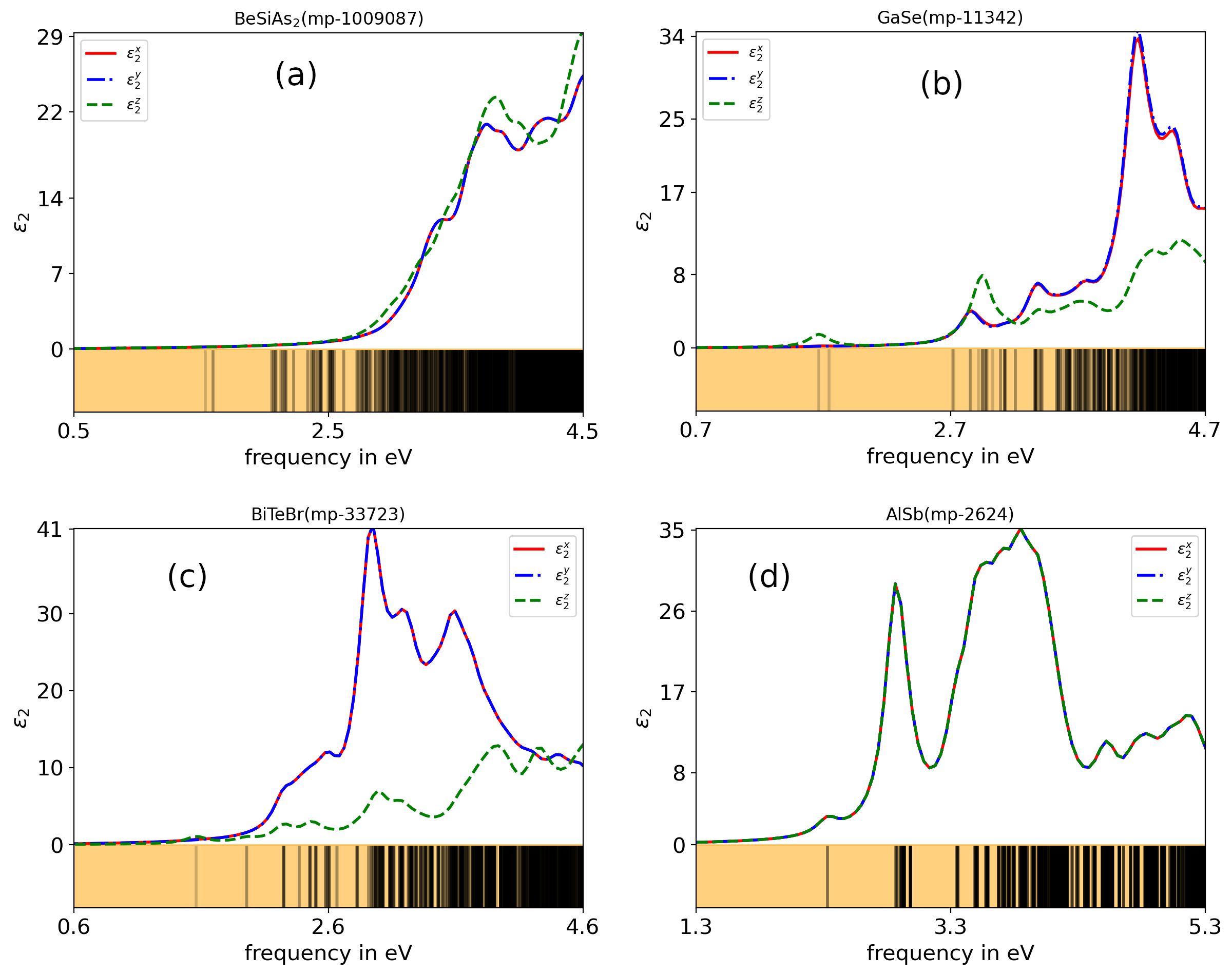}
\caption{Imaginary part of the dielectric function for light polarization along
	$x$ (red), $y$ (blue) and $z$ (green) axis obtained from G$_0$W$_0$-BSE
	calculation for (a) BeSiAs$_2$ (mp-1009087), (b) GaSe (mp-11342), (c)
	BiTeBr (mp-33723), and (d) AlSb (mp-2624). The yellow panel on the bottom of each figure shows the excitation energies with vertical
	lines.} 
\label{fig4}
\end{figure}

Fig. ~\ref{fig3} shows the $\alpha_\mathrm{int}$ computed for the case of light
polarization along $x$ (red bars), $y$ (blue bars), and $z$ (green bars) axis for all 48 materials under consideration. We find that among materials with low EBE ($<$ 200 meV) 14 materials that have a $\alpha^\mathrm{avg}_\mathrm{int}$ values larger than the cutoff of $3.6 \times10^4$ cm$^{-1}$ eV. 

%Materials such as BeSiAs$_2$ (mp-1009087) and GaSe
%(mp-11342), which have an optical gap of 1.54 eV and 1.67 eV respectively, do not overcome this cutoff. 

In Fig.~\ref{fig4} we show the absorption spectra of
four representative materials, BeSiAs$_2$ (mp-1009087), GaSe (mp-11342), BiTeBr (mp-33723) and AlSb (mp-2634). The excitation energies for each of these materials are shown as vertical lines in the yellow panels underneath their
spectra. 

Let us examine the case of BeSiAs$_2$, Fig. ~\ref{fig4}a. For this material, we can see the lowest energy exciton is at 1.54 eV, its optical gap, and there are quite a few excitons in the 2-2.5 eV energy range. However, most of these excitons are `dark', i.e. they do not yield an appreciable $\alpha_\mathrm{int}$. Most of the `bright' excitons for
BeSiAs$_2$ lie above $\sim$ 3 eV and thus this material has a low
$\alpha_\mathrm{int}$ value in the visible light region. Hence it does not meet the cutoff for criteria 2. 

In the case of GaSe, which has a 1.67 eV optical gap, there are very few bright excitons below 2 eV and some bright excitons
above 2.7 eV. The low number of excitons results in a poor $\alpha_\mathrm{int}$ value (Fig.
~\ref{fig4}b). Thus, this material also does not meet the criteria 2 cutoff.

Transition metal dichalcogenides MoSe$_2$ and WSe$_2$ and other layered material such as
BiTeBr (Fig.~\ref{fig4}c) absorb significantly poorly if the light polarization is along the z-axis ($\sim$ 30 $\%$ compared to light polarization
along x or y). This results in a low $\alpha^\mathrm{aniso}_\mathrm{int}$ for many of the layered materials. For example, GaTeCl and CsYZnSe$_3$ which are also layered materials show significantly lower absorption if the light polarization is along the axis with the vdW bonding. In the case of GaTeCl (CsYZnSe$_3$) we see 66 (79) $\%$ and 92 (38) $\%$ decrease in total visible light absorption if the light polarization is along the y and z-axis respectively compared to if the polarization is along the x axis. Ga$_2$Se$_3$, RbInTe$_2$, Sr(InTe$_2$)$_2$, ZnTe, Ta$_2$SnO$_6$ are some of the
other materials which have a preference in absorbing visible lights of certain polarization than the others.  

The absorption spectra of
one of the gifted materials AlSb (mp-2634), which satisfies both the criteria 2 and 3 has been shown in Fig.~\ref{fig4}d. It has an EBE of 110 meV, $\alpha^\mathrm{avg}_\mathrm{int}$ of  7 $\times10^4$ cm$^{-1}$ eV, and $\alpha^\mathrm{aniso}_\mathrm{int}$  of 1. 

The absorption spectra of all the 48 materials can be found in section VI of the Supplementary Information. Supplementary Table S4 lists the $\alpha^\mathrm{aniso}_\mathrm{int}$, $\alpha^\mathrm{avg}_\mathrm{int}$, $\alpha^{x}_\mathrm{int}$, $\alpha^{y}_\mathrm{int}$, and $ \alpha^{z}_\mathrm{int}$ in the visible-light (1.7--3.5 eV) range of all the 48 materials. 

\begin{figure}[h!]
\centering
\includegraphics[width=\linewidth]{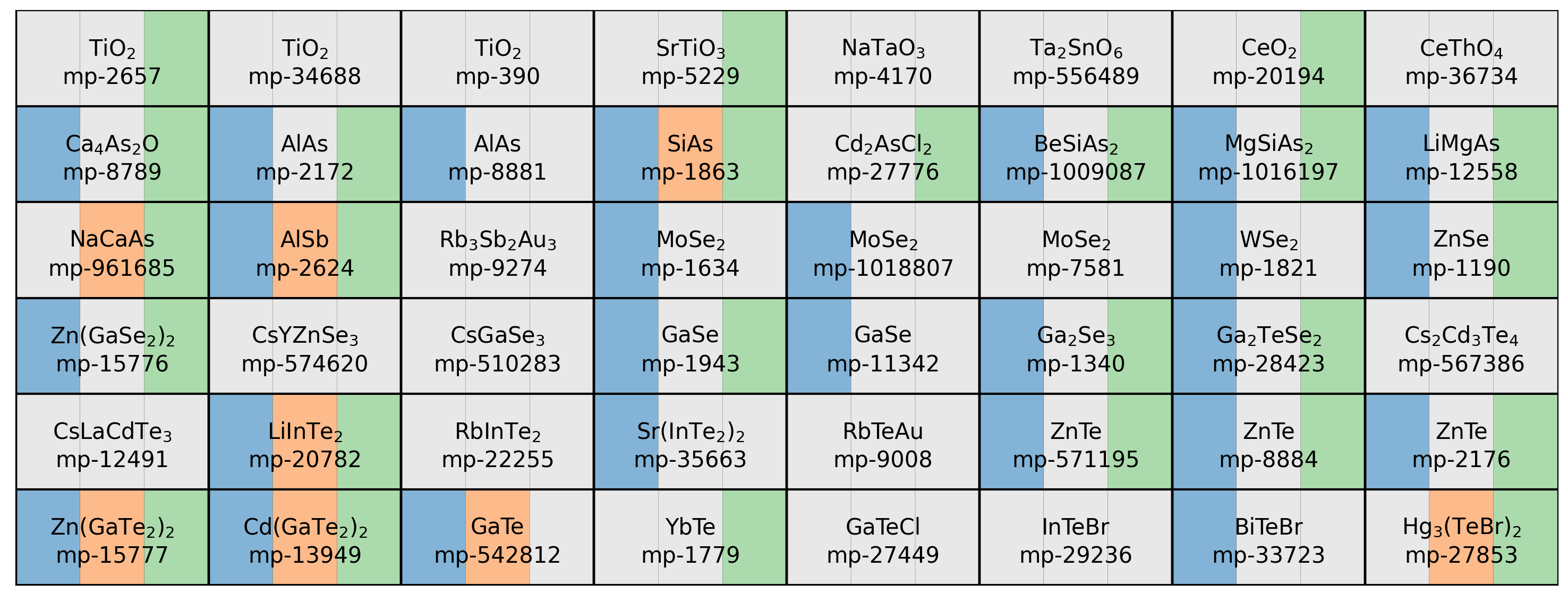}
	\caption{The figure shows a color-coded map that indicates which of the screening criteria are satisfied by the 48 materials. The materials that satisfy the EBE criteria 1 are shaded in blue. The materials that satisfy both criteria 2 and 3 of meeting the cutoff of $\alpha^\mathrm{aniso}_\mathrm{avg}$ and $\alpha^\mathrm{aniso}_\mathrm{int}$ in the visible light region (1.7--3.5 eV) are shaded in orange. The materials that satisfy criteria 2 and 3 in the UV region (3.5--4.2 eV) are shaded in green.}
\label{fig5}
\end{figure}

In Figure~\ref{fig5} we show a color-coded map to indicate which of the screening criteria are satisfied by the 48 materials. Materials that satisfy the EBE criteria are shaded in blue. Materials that satisfy both criteria 2 and 3 of meeting the cutoff of $\alpha^\mathrm{aniso}_\mathrm{avg}$ and $\alpha^\mathrm{aniso}_\mathrm{int}$ in the visible light region (1.7--3.5 eV) are shaded in orange. Materials that satisfy criteria 2 and 3 in the UV region (3.5--4.2 eV) are shaded in green. In this color-coded map, we can see that only 6 materials meet the three criteria for visible light photocatalysis. These 6 materials are AlSb, SiAs, GaTe, Zn(GaTe$_2$)$_2$, Cd(GaTe$_2$)$_2$, and LiInTe$_2$. All of these materials have been previously synthesized.\cite{singh2019robust} SiAs and GaTe are layered materials. These layered materials can be extracted in two-dimensional forms, which are known to offer high specific surface area, tunability, and potential for heterostructuring.~\cite{torrisi2020two, singh2014computational, paul2017computational} In a recent computational study, Torrissi et al. have shown the two-dimensional SiAs is selective for CO production and has good absorption in the visible light region.~\cite{torrisi2020two} On the other hand, two-dimensional GaTe has been shown to exhibit a sharp decrease in photoluminescence during the transition from the many to few-layer limit.~\cite{torrisi2020two}

The materials that satisfy all the criteria in the UV region are markedly more, total 18. In Figure~\ref{fig5} these materials are shaded in blue and green colors. We observe that most of the oxides do not make the list but several arsenides, tellurides, and selenides emerge as suitable candidates.  Supplementary Table S5 lists the $\alpha^\mathrm{aniso}_\mathrm{int}$, $\alpha^\mathrm{avg}_\mathrm{int}$, $\alpha^{x}_\mathrm{int}$, $\alpha^{y}_\mathrm{int}$, and $ \alpha^{z}_\mathrm{int}$ in the UV-light range (3.5-4.2 eV) of all the 48 materials. 

\subsection*{Wannier-Mott Model vs BSE}
In the case of most inorganic semiconductors, excitons can be approximated as an electron in the conduction band that is bound with a hole in the valence band through some form of screened Coulomb interaction and can be described using the
Wannier-Mott (WM) model. This model uses effective mass approximation which describes electrons and holes as free particles having a parabolic dispersion and characterized by effective masses that depend on the crystal structure. 

Within this approximation, the wave-function of relative electron-hole motion or exciton can be found from the Schrodinger equation very similar to the one describing the electron state in a hydrogen atom where we replace the electron mass, $m_0$, with reduced effective mass, $\mu=\frac {m_e m_h} {m_e+m_h}$, and replace the unscreened Coulomb interaction, $\frac{e^2} {r}$, between the
electron-hole pair with the screened one,  $\frac{e^2} {\epsilon r}$. Where $m_e$ ($m_h$) is the electron (hole) effective masses and $\epsilon$ is the dielectric constant of the material. By solving the hydrogen atom like Schrodinger's equation we find the binding energy of the ground excited state
(EBE$_\mathrm{WM}$),

\begin{equation}
    \mathrm{EBE}_\mathrm{WM}=\frac {\mu e^4} {2 \hbar^2 \epsilon^2}
\end{equation}

Computationally it is much less expensive to obtain EBE using the WM model than solving the BSE. The input parameters in WM model are the dielectric constant and the effective masses, which can be obtained with reasonable accuracy using less the expensive DFT
calculations \cite{gajdovs2006linear}. Here, we test the applicability of WM model for the materials under consideration and the possibility of using WM instead of the expensive BSE calculations for future solar-materials discovery.

 Note that the WM approximation is valid when the interaction potential is slowly varying over the dimension of an unit cell or in other words the dimension of the lowest energy exciton is much larger than the lattice dimension. In this case we get an EBE of the order of ~0.2 eV. The EBE of the majority of the materials considered in this work is in the vicinity of 0.2 eV value, and the range of the EBE is 0.11--0.59 eV. 

To find the degree of correlation between the EBE computed using BSE and WM model (EBE$_\mathrm{WM}$), we calculate the Pearson's correlation coefficient (PCC) and the Spearman’s correlation coefficient (SCC) \cite{weaver2017introduction}. PCC indicates the strength of linear relationship between two variables whereas, SCC doesn't assume any particular functional dependence to quantify the strength of the correlation. We find a PCC value of 0.77 and SCC value of 0.76 between the BSE EBE and the WM EBE, which suggests a strong correlation between the two. Therefore, it would be quite reasonable to use WM model for a computational data-driven discovery of materials that have low EBE instead of using the computationally much more expensive BSE calculations. 

The high PCC value also suggests that we can scale the EBE obtained from WM model by a factor of 0.17 (see Supplementary Information Table S5)  to get a reasonable prediction for the value obtained by solving the BSE. We find that such a scaling of the WM EBE to obtain the BSE EBE results in a mean squared error of 0.11 eV and root mean squared deviation of 0.12 eV, which is similar to the accuracy of standard GW-BSE calculation. 

\section*{Summary and Discussion}
In summary, we have explored excitonic effects in the optical properties of 48
promising CO$_2$ reduction photocatalysts. We have designed a screening strategy considering the optical properties of materials that are crucial for such photocatalysts, i.e, the exciton binding energies, the capability to absorb visible light, and the degree of anisotropy in the absorption of light with different polarizations. 

Using this screening strategy we have identified 6 materials that are suitable
for visible light photocatalysis. Interestingly, to our knowledge, none of
these materials have been extensively studied to be used as photocatalysts for
CO$_2$ reduction. Three of these materials have been studied before using
various theoretical and experimental methodology, i.e, SiAs
\cite{wang2019class,kunioka1973optical}, AlSb
\cite{wing2019comparing,Blunt1954}, and GaTe \cite{antonius2018orbital}.
However, the other three materials LiInTe$_2$, Zn(GaTe$_2$)$_2$ and
Cd(GaTe$_2$)$_2$ have remained largely unexplored. It is noteworthy that all of
these materials have been previously experimentally synthesized
\cite{Wadsten1965,Blunt1954,JulienPouzol1979,Kuehn1985,Errandonea2013,Neumann1993}.
Our screening strategy also identifies 18 materials that are suitable for UV
light photocatalysis. 

Using the relatively large amount of the high-quality BSE EBE data generated in this work, we show that the Wannier-Mott model can be used for estimating excitonic binding energies in a computationally cheaper manner for large-scale computational material discovery of photoabsorber materials without the need of performing the extraordinarily expensive BSE calculations. 

Overall, our work presents a new paradigm towards the inclusion of optical properties for data-driven discovery of materials for solar energy conversion applications.

\section*{Methods}

All the first-principles calculations reported in this study were performed
using Vienna Ab initio Simulation Package (VASP) \cite{kresse1999ultrasoft,shishkin2006implementation,sander2015beyond}. The
Kohn-Sham wavefunctions and eigenvalues were obtained using a plane wave basis with an energy cutoff of 520 eV. The exchange correlation functional was approximated using Generalized Gradient Approximation
\cite{perdew1996generalized}. The Brillouin zone was sampled using a $k$-grid with 100 kpoints per \AA$^{-3}$ of the reciprocal cell. The crystal structures were obtained by relaxing the atomic positions and the lattice parameters by a energy convergence cutoff of 0.0005 eV $\times$ number of atoms in the cell, which has been found to produce well-converged structures in most instances
\cite{jain2011high}. 

The macroscopic dielectric constants reported here were
calculated using density functional perturbation theory following the
formalism originally introduced by Baroni et. al. \cite{baroni1986ab} and later
implemented for projector-augmented wave methodology by Gajdo{\v{s}} et.
al \cite{gajdovs2006linear}. The electron and hole effective masses were computed using parabolic line fitting at the direct band edges. We used the SUMO code \cite{ganose2018sumo} to perform the effective mass calculations.     

The quasiparticle energies were obtained by using many body perturbations theory within both G$_0$W$_0$ and GW$_0$ approximation for the self-energy
operator \cite{hedin1965new,shishkin2006implementation,shishkin2007self}. The
basis set size for the response functions and W was chosen to include all
plane waves up to an energy cutoff of 340 eV. The number of unoccupied bands
included in the GW calculation was selected such that the QP gap is converged
to within 0.1 eV. The excitonic effects were studied by solving Bethe-Salpeter
equation within Tamm-Dancoff approximation
\cite{rohlfing2000electron,sander2015beyond}. We included all the
occupied/unoccupied bands which were within 1.5 eV from band edges in the BSE
calculation to ensure convergence of the absorption spectra. However, the
convergence of the BSE absorption spectra with the number of $k$-points
($N^\mathrm{BSE}_k$) is often quite challenging and computationally expensive
\cite{sander2015beyond}. Nonetheless, in this study we selected the $N^\mathrm{BSE}_k$
such that the EBE converged to within 0.1 eV. See supplementary Table. S6 for a list of all the converged parameters for each of the 48 materials.

We have developed a Python code for performing automated first-principles calculations within GW-BSE framework and used it carry out all the calculations reported in this article. Our code is built upon open-source Python packages developed by the Materials Project, such as pymatgen \cite{mostofi2008wannier90}, fireWorks \cite{jain2015fireworks}, and atomate \cite{mathew2017atomate} to achieve complete automation of the entire multi-step GW-BSE computational framework that requires several convergence parameters. GW-BSE calculation is extremly sensitive to multiple interdependent convergence parameters such as number of bands ($N_\mathrm{b}$) included in the GW self energy calculation, number of k-points (N$^{BSE}_k$) used to sample the Brillouin zone in the BSE calculation etc. Our workflow is capable of performing automated convergence test of all these parameters and using those values to carry out fully converged GW-BSE calculation. However, due to limitation of the computational resources available, we have restricted ourselves with converging $N_\mathrm{b}$ and $N^\mathrm{BSE}_k$, two most sensitive parameters in the GW-BSE calculation. For several other convergence parameters such as, number of (imaginary) frequency grid points in GW calculation, energy cutoff for response function we have chosen a value large enough to produce quasiparticle energies within 0.1 eV. To reduce the computational cost associated with performing convergence tests we have judiciously used COHSEX and DFT level calculations which are significantly cheaper than full GW-BSE calculation. Details about these choices to perform efficient convergence calculations for GW-BSE calculations will be included in a forthcoming manuscript. In this workflow, we have also employed Wannier90 \cite{mostofi2008wannier90}, a program for calculating maximally-localized Wannier functions to perform the interpolation required to obtain quasiparticle bandstructure. 

\bibliography{excitonic}

\section*{Data availability}
All data supporting the findings of this work are available in the article and its Supplementary Information. Extra data and machine readable data are available upon reasonable request to the authors.

%\section*{Supplementary information}

\section*{Contributions}
T.B performed the simulations and calculations. T.B. and A.K.S. analyzed the data and designed the research. Both the authors contributed to designing the research methods, interpreting the data, and writing the manuscript.

\section*{Acknowledgements}
This work was supported by the Arizona State University start-up funds and in part as part of ULTRA, an Energy Frontier Research Center funded by the U.S. Department of Energy (DOE), Office of Science, Basic Energy Sciences (BES), under Award \# DE-SC0021230 (GW-BSE high-throughput simulations). In addition, Singh acknowledges support by the NSF DMR-grant NSF-DMR \#1906030. The authors acknowledge the San Diego Supercomputer Center under the NSF-XSEDE Award No. DMR150006 and the Research Computing at Arizona State University for providing HPC resources. This research used resources of the National Energy Research Scientific Computing Center, a DOE Office of Science User Facility
supported by the Office of Science of the U.S. Department of Energy under Contract No. DE-AC02-05CH11231.

\end{document}